\documentclass[journal]{vgtc} 
\usepackage{cite}
\usepackage{xcolor}
\usepackage{graphicx}
\usepackage{caption}
\usepackage[table]{xcolor}
\usepackage{pgfplotstable}
\pgfplotsset{compat=1.18}

\pgfplotsset{
  colormap={redgreen}{
    rgb(0cm)=(1,0.9,0.9)   % light red
    rgb(0.5cm)=(1,1,1)     % white
    rgb(1cm)=(0.85,1,0.85) % light green
  }
}
\usepackage[most]{tcolorbox}
\usepackage{soul}

\usepackage[usenames, dvipsnames]{xcolor }            
\usepackage{xcolor}       % In preamble

\newif\ifuseerrorcolors
\useerrorcolorstrue  % set to \useerrorcolorstrue to enable again
\ifuseerrorcolors
    \definecolor{errREASON}{HTML}{2E86AB}
    \definecolor{errTrans}{HTML}{F28E2B}
    \definecolor{errCOHERENCE}{HTML}{59A14F}
    \definecolor{errMETRIC}{HTML}{E15759}
    \definecolor{errPERPCEPTION}{HTML}{AA8CFF}
\else
  \colorlet{errREASON}{.}
  \colorlet{errTrans}{.}
  \colorlet{errCOHERENCE}{.}
  \colorlet{errMETRIC}{.}
  \colorlet{errPERPCEPTION}{.}
  \newcommand{\errchip}[2]{#2}

\fi

\newcommand{\errchip}[2]{%
  \begingroup
    \fboxsep=0.6ex \fboxrule=0pt
    \colorbox{#1!12}{\textcolor{#1}{\sffamily\bfseries\footnotesize #2}}
  \endgroup
}
\onlineid{0}

\vgtccategory{Empirical Studies}

\vgtcpapertype{Evaluation}

\title{Do MLLMs See What We See? \\ Analyzing Visualization Literacy Barriers in AI Systems}

\author{%
  Mengli (Dawn) Duan\textsuperscript{*}, 
  Yuhe (Sissi) Jiang\textsuperscript{*},
  Matthew Varona, 
  and Carolina Nobre
}

\authorfooter{
  \item
   \thanks{\textsuperscript{*}Equal contribution.}
   \item
    Mengli (Dawn) Duan, 
  Yuhe (Sissi) Jiang,
  Matthew Varona, 
  Carolina Nobre are with 
  	University of Toronto, Canada.
  	E-mail: \{mengli.duan, yuhe.jiang, \}@mail.utoronto.ca, \{varona, cnobre\}@cs.toronto.edu
   
}

\abstract{Multimodal Large Language Models (MLLMs) are increasingly used to interpret visualizations, yet little is known about \emph{why} they fail. We present the first systematic analysis of barriers to visualization literacy in MLLMs. Using the regenerated Visualization Literacy Assessment Test (reVLAT) benchmark with synthetic data, we open-coded 309 erroneous responses from four state-of-the-art models with a barrier-centric strategy adapted from human visualization literacy research.
Our analysis yields a taxonomy of MLLM failures, revealing two machine-specific barriers that extend prior human-participation frameworks. Results show that models perform well on simple charts but struggle with color-intensive, segment-based visualizations, often failing to form consistent comparative reasoning. Our findings inform future evaluation and design of reliable AI-driven visualization assistants.
}

\keywords{Visualization Literacy, Multimodal Large Language Model, Evaluation Study}

\teaser{
  \centering
  \includegraphics[width=\linewidth, alt={placeholder}]{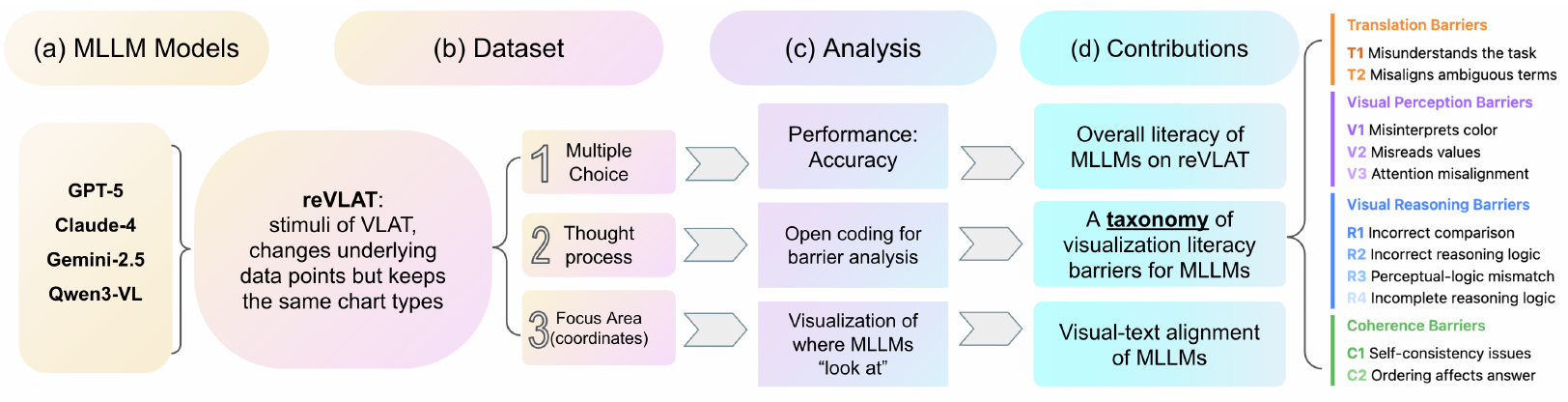}
  \caption{%
  	An overview of the proposed framework for identifying barriers in visualization literacy of MLLMs including (a) SOTA MLLM Models evaluated, (b) the dataset reVLAT: an alternative version of VLAT with synthetic underlying data, (c) MLLM responses with reasoning and points of focus collected for analysis, (d) insights into MLLM rationale drawn from the results, with an overview of the categorized barriers for MLLMs. %
  }
  \label{fig:teaser}
}

\graphicspath{{figs/}{figures/}{pictures/}{images/}{./}} % where to search for the images

\usepackage{tabu}                      % only used for the table example
\usepackage{booktabs}                  % only used for the table example
\usepackage{lipsum}                    % used to generate placeholder text
\usepackage{mwe}                       % used to generate placeholder figures

\usepackage{mathptmx}                  % use matching math font

\begin{document}
%% The ``\maketitle'' command must be the first command after the
%% ``\begin{document}'' command. It prepares and prints the title block.
%% the only exception to this rule is the \firstsection command
% \firstsection{Introduction}

\maketitle

\section{Introduction}
%% \section{Introduction} %for journal use above \firstsection{..} instead

Data visualization plays a fundamental role in modern data analysis and communication, serving both to transform raw data into interpretable insights and to facilitate communication between researchers and their readers. Visualization literacy, known as the ability to read, interpret, and reason about visual data representations, has been extensively studied across diverse populations~\cite{grammel2010How,alper2017elementary,statton2022proliferation}. However, with recent advances in AI systems, a new category of \textit{readers} has emerged: Multimodal Large Language Models (MLLMs) such as~\emph{GPT} and ~\emph{Claude}, which are capable of reading both text inputs and image inputs. 

These systems are increasingly being deployed as assistants in professional, educational, and general analytical contexts to extract data, synthesize information, and interpret complex visualizations. Recent work has conducted preliminary studies on MLLM  capabilities in basic visual tasks, showing that they have competitive yet imperfect performance in understanding data visualizations~\cite{bendeck2024empirical, li2024visualization,hong2025llms, li2025see}. As these models are highly complex and opaque, understanding their visualization literacy capabilities and limitations becomes critical to enhance human-machine collaboration and trustworthiness.

A recent work by~\cite{nobre2024reading} established a validated taxonomy to analyze human visualization literacy barriers, providing a foundation for understanding the mental gaps that lead to incorrect interpretation among human participants. However, despite some failure patterns noted in recent works~\cite{li2024visualization,li2025see}, no equivalent systematic framework exists for MLLMs. This raises a natural question: \textbf{do MLLMs encounter similar barriers to those faced by humans, or do they exhibit distinct failure patterns due to their fundamentally different perceptual and reasoning mechanisms?} In this work, we conduct a comprehensive study on why and how MLLMs fail in visualization interpretation tasks, and extend the taxonomy developed in~\cite{nobre2024reading} to categorize common visualization literacy barriers for MLLMs.

To address this gap, we conduct a comprehensive study on MLLM failure cases in visualization interpretation tasks. We adapt the barrier-centric, open-coding methodology inspired by~\cite{nobre2024reading} and apply it to MLLMs on a Visualization Literacy Assessment Test (VLAT) variant with alternate data values to minimize data leakage, $\text{reVLAT}$~\cite{hong2025llms}. We collect and analyze reasoning traces from four state-of-the-art (SOTA) MLLMs, allowing us to derive an extended taxonomy that categorizes common MLLM failures.
The main contributions of our work are (a) an overall visualization literacy assessment of SOTA MLLMs on reVLAT, offering insights into MLLM capabilities in visual tasks, and (b) an extended taxonomy of visualization literacy barriers for MLLMs, categorizing MLLM failures in visualization understanding.
More specifically, we show that MLLMs exhibit errors similar to humans, categorized under~\errchip{errTrans}{Translation Barriers} and~\errchip{errPERPCEPTION}{Visual Perception Barriers}, while also presenting machine-specific failures, categorized under~\errchip{errREASON}{Visual Reasoning Barriers} and {\errchip{errCOHERENCE}{Coherence Barriers}.

\section{Related Work}

This work investigates literacy barriers to visualization in MLLMs by bridging established frameworks for human participants and recent evaluation pipelines for (M)LLMs\footnote{LLMs and Multimodal LLMs}. In this section, we review (1) visualization literacy assessments and barrier analysis developed for human participants, (2) empirical evaluations of (M)LLM performance on visualization tasks, and (3) emerging efforts to characterize (M)LLM failure modes.

\subsection{Visualization Literacy Assessment}
Visualization literacy refers to \textit{the ability to read and interpret data visualizations}~\cite{VisLiteracyMeaning}, which has been extensively studied in the literature through structured assessments. Early and influential work includes the VLAT by~\cite{vlat}, comprising 12 chart types and 53 multiple-choice questions, which laid the foundation for standardized testing of human participants' visualization literacy. Following VLAT, several other benchmarks are constructed focusing on efficiency or critical thinking, such as Mini-VLAT~\cite{pandey2023mini} and CALVI~\cite{ge2023calvi}. 

Beyond quantitative evaluations, exploring failure cases in visual tasks is also a central focus in academic research. A recent key work extended VLAT questions to an open-response format to enable qualitative analysis of participants' internal reasoning processes~\cite{nobre2024reading}. Through open coding of the incorrect answers collected from 120 participants, they derived a complete taxonomy of human visualization literacy barriers, categorizing conceptual and operational gaps.

Inspired by the success of this work, we adapt and extend this barrier-centric approach to AI systems, aiming to provide a comprehensive and systematic understanding of AI failure patterns.

\subsection{AI Systems Capability in Visualization Tasks}

Recent evolution of AI systems, especially LLMs and Multimodal LLMs (MLLMs)\footnote{We focus on MLLMs with~\textbf{image} and text as the two input modalities.}, has made them increasingly deployed as assistants to interpret and interact with data visualizations~\cite{choe2024enhacing}. Therefore, it is critical to evaluate their visualization literacy.

Existing work primarily focused on empirical evaluation of LLMs and MLLMs using accuracy as the main metric.~\cite{bendeck2024empirical} reported that early LLMs achieved 87\% on Chart QA~\cite{CQA2020Kim} when the underlying data was provided alongside the visualization, but performance decreased to 31\% with the underlying data removed.~\cite{li2024visualization} conducted a comparative study of MLLMs, showing that these models achieved near-human performance on VLAT (58\% by the models v.s. 66\% by humans). However, these results raised concerns about \textit{data leakage}, as open datasets such as VLAT are highly likely to be included in the training corpus of (M)LLMs. To mitigate this risk,~\cite{hong2025llms} introduced reVLAT, a benchmark that regenerates VLAT visualizations with a new set of synthetic data while keeping the same chart types and questions. Their findings revealed a significant drop in accuracy for MLLMs, suggesting potential flaws in their understanding of common visualizations.

While these works established MLLMs' basic capabilities in visual tasks, they often lack a complete analysis of \textit{why the models fail}. Our work, inspired by~\cite{nobre2024reading} and leveraging the reVLAT pipeline, directly addresses this gap by shifting the focus from accuracy measurements to qualitative analysis of MLLM reasoning rationales, providing a deeper understanding of MLLM visualization literacy barriers.

\subsection{Understanding MLLM Failure Modes}

In the aforementioned works, several potential causes of MLLM failure are discussed. For example, value retrieval is noted as one of the main sources of error~\cite{li2024visualization}, where MLLMs struggle to extract precise values from visualizations. Specific chart types that involve complex spatial or perceptual reasoning could also limit models' performance~\cite{li2024visualization,hong2025llms}. Following the development of reVLAT,~\cite{li2025see} conducted a closer study of the recall problem, highlighting that MLLMs may consistently output answers based on memory or bias even when the image is missing. 

Despite these explorations into isolated issues, no existing work provides a comprehensive categorization of MLLM visualization literacy barriers across multiple SOTA models. Our contribution is to fill this gap by systematically analyzing a broad set of SOTA MLLMs on reVLAT, deriving a comprehensive failure taxonomy, and investigating whether MLLMs share the same barriers as humans or possess distinct, machine-specific rationales.

\section{Methodology}

In this section, we present the experimental design, the configurations for MLLM response collection, and the analytical plan used to derive the MLLM visualization literacy barrier taxonomy.

\subsection{Study Design}

We adapted the evaluation pipeline from~\cite{hong2025llms} to collect MLLM responses for a qualitative analysis inspired by~\cite{nobre2024reading}.\newline

\noindent{\textbf{Models.}} As shown in Figure~\ref{fig:teaser}, four SOTA MLLMs are selected to ensure comprehensive insight into contemporary failure patterns, including~\emph{GPT-5 (GPT)}~\cite{openai_2025}, \emph{Claude-Sonnet-4 (Claude)}~\cite{anthropic_2025a}, \emph{Gemini 2.5}~\cite{google_gemini_2025}, and \emph{Qwen3-VL (Qwen)}~\cite{qwen_2025}. All models are accessed through the OpenRouter API~\cite{openrouter}; and can take images as direct input. \newline 

\noindent{\textbf{Dataset.}} We leverage the existing reVLAT Github repository\cite{hong2025llms} for data generation and MLLM response collection. reVLAT consists of 53 questions across 12 chart types, where each chart is regenerated with a given random seed that creates a synthetic underlying data distribution (different from the original VLAT) to mitigate the data leakage concern. All images are high-resolution (1233 × 600 pixels). Subsequently, questions and answer options are reformatted into a JSON schema and stored locally for better reproducibility. Each question is then linked to its corresponding visualization stimulus. Color palettes and font sizes are standardized to replicate the controlled perceptual conditions of the original VLAT. For evaluation, the dataset is serialized into JSON files compatible with MLLM prompting. The entire pipeline is implemented in Python; a global random seed is used to control the regenerated data. We fixed the random seed for better consistency and reproducibility.\newline

\noindent{\textbf{Response Augmentation.}} Inspired by~\cite{nobre2024reading}, we augmented the standard multiple-choice question-answering pipeline to collect three distinct response components from MLLMs, which enables further qualitative analysis on MLLMs' reasoning rationales:

\begin{itemize}
    \item In addition to the original answer choices, we add a~\texttt{"I'm not sure"} choice to allow the model to explicitly admit its incapability to answer the question rather than forcing a guess.

    \item In addition to the selected choice, we ask the model to output its rationales supporting its selected choice with up to 5 sentences.

    \item Lastly, we ask the model to output a set of (x,y) coordinates (up to 5) that highlight its points of focus (i.e. where it~\texttt{"looks at"}) during the question answering phase\footnote{We replaced direct \textit{sketches} on the images collected in~\cite{statton2022proliferation} by points of focus because MLLMs could not stably output high-quality images.}.
\end{itemize}

Following~\cite{hong2025llms}, MLLMs are minimally prompted to output these three types of responses in a structured JSON format (see Appendix~\ref{sec:prompts_system} for the detailed example prompt).

\subsection{Pilot Experiments}

We conducted a set of pilot experiments with Claude 4 on a random sample of 20 questions from the 53 in reVLAT and collected insights that are incorporated into the full evaluation.\newline

\noindent{\textbf{MLLM Temperature.}} The temperature parameter ($T$) governs the variance of MLLM responses: higher $T$ increases the diversity and randomness of the generated text, while lower $T$ makes the output more deterministic~\cite{li2025temperature}. In the pilot experiments, we tested MLLMs with $T=0.0,0.5,1.0$, and observed no significant change in overall performance accuracy (around 79\%). Therefore, we set $T=0.0$ for the full experiment for better reproducibility.\newline

\noindent{\textbf{Choice Ordering.}} Prior work~\cite{hong2025llms} showed that choice orderings have a non-negligible effect on MLLM responses. We observed a similar trend in the pilot experiment, where evaluations across different choice orderings yielded an overall performance ranging from 67\% to 79\%. This leads us to sample multiple orderings in the full experiment to minimize possible side effects introduced by choice orderings. 

\subsection{Full Experiment}

Based on the insights from the pilot experiments, we conducted the full experiment by generating all possible choice orderings for each question, then randomly sampling 5 orderings without replacement to evaluate the MLLMs. Each question and ordering pair was run in a separate session to prevent the model from carrying over context or memory across trials.
In total, 1060 trials were conducted (53 questions $\times$ 5 orderings $\times$ 4 models), resulting in 265 trials per model. The computational cost totaled 3.86 million tokens, which cost 11.79 USD.

\subsection{Multi-level Analysis} We conducted a multi-level analysis of the collected responses:
\begin{enumerate}
    \item \textit{Performance analysis}: calculation of overall accuracy metrics across chart types and model families;
    \item \textit{Open coding}: following the methodology of~\cite{nobre2024reading}, we systematically apply open coding to the reasoning traces extracted to establish an extended taxonomy of MLLM visualization literacy barriers (309 incorrect answers in total across four models);
    \item \textit{Quantitative analysis}: numerically visualize the coded barrier distributions at the model-level and chart-level.
\end{enumerate}

\section{MLLM Visualization Literacy Barriers}\label{sec:barrier}

In this section, we present a taxonomy of visualization literacy barriers observed in MLLMs, developed through open coding of experimental responses. Consistent with prior work on human visualization literacy~\cite{nobre2024reading}, our analysis reveals barriers at multiple stages of visualization interpretation, ranging from translating a textual query to extracting and reasoning over visual information. In contrast to human readers, MLLMs exhibit distinctive reasoning and coherence failures attributable that arise from their tokenized generation process~\cite{wang2023selfconsistencyimproveschainthought, wei2023chainofthoughtpromptingelicitsreasoning}, underscoring the importance of characterizing these barriers.

\subsection{\errchip{errTrans}{Translation Barriers}}

Translation barriers arise during the initial interpretation of the question, when the model is instructed to infer the task requirements and identify the relevant elements in the visualization. As such, barriers are conceptual in nature and can lead to incorrect value extraction or misaligned reasoning. In our taxonomy, we identify two primary forms of translation failures: \textit{\textbf{T1 "Misunderstands the task"}} and \textit{\textbf{T2 "Misaligns ambiguous terms"}}.

\noindent{\errchip{errTrans}{T1: Misunderstands the task}}
The model fails to correctly identify the intended visual analytic task. For instance, when asked to determine a range, the MLLM may instead compute a difference between two individual values or interpret the question as seeking a trend. Similar to human participants~\cite{nobre2024reading}, misunderstanding the task results in downstream errors, like retrieving the wrong visual element or drawing an incorrect conclusion.

\noindent{\errchip{errTrans}{T2: Misaligns ambiguous terms}}
This barrier reflects challenges related to language alignment or semantic ambiguity, in which terms such as \texttt{"large"} and \texttt{"small"} are interpreted differently by the model than intended in the question. The model may struggle to apply consistent definitions or to adopt contextual interpretations of descriptive language. Furthermore, unclear or underspecified instructions can trigger inference errors, leading the model to focus on irrelevant chart elements or categories.

%%%%%%%%%%%%%%%%%%%%%%

\subsection{\errchip{errPERPCEPTION}{Visual Perception Barriers}}

Visual perception barriers arise when the model struggles to correctly extract low-level visual information from the chart, such as colors, values, or specific data marks. Similar to the encoding-related issues observed in human readers~\cite{nobre2024reading}, these barriers reflect difficulties in interpreting perceptual cues rather than misunderstanding the task itself. Errors at this stage often propagate into subsequent reasoning steps, leading to incorrect comparisons or conclusions, even though the overall task interpretation is correct. We therefore identify three common categories of perceptual failures: \textit{\textbf{V1 "Misinterprets color"}} , \textit{\textbf{V2  "Misreads values"}} and \textit{\textbf{V3 "Attention misalignment"}}.

% \begin{itemize}
%     \item V1: Misinterprets color
%     \item V2: Misreads values
%     \item V3: Attention misalignment
% \end{itemize}

\noindent{\errchip{errPERPCEPTION}{V1: Misinterprets color}}
Color misinterpretation occurs when the model misreads the color encoding in the visualization. This error occurs particularly for visualizations that use gradients or quantized color scales, such as choropleth maps. The model may correctly locate the target region but misinterpret the corresponding legend interval (e.g., inferring \texttt{4 to 6\%} instead of \texttt{6 to 8\%}). As a result, it reads the wrong value from the colormap.

\noindent{\errchip{errPERPCEPTION}{V2: Misreads values}}
Value misreading occurs when the model incorrectly extracts a numerical value from the visualization. We observe two variants of this barrier, in combination with direct and indirect wrong readings. Direct wrong reading arises when the model explicitly states a value that does not correspond to the visual evidence, such as reporting \texttt{40} when the plotted value aligns with \texttt{50} on the y-axis. In contrast, indirect wrong reading occurs when the model does not explicitly state a misread value, but its subsequent reasoning implies an incorrect perception. For example, an incorrect comparison or conclusion may reveal that the model internally misread a value, even if the erroneous reading is not directly articulated. Hence, the error refers to reading the correct data element but misinterpreting its numeric scale or axis alignment.

\noindent{\errchip{errPERPCEPTION}{V3: Attention misalignment}}
Attention misalignment occurs when a model's “attention focus” is on an incorrect visual region or element, despite understanding the task. The model may identify the wrong bar, segment, or point due to misplaced “attention focus” with an offset or a lack of precision in visual grounding. In some instances, this appears as uncertainty (e.g., returning \texttt{"not sure"}) when the wrong area of “attention focus” is extracted. It therefore suggests an inadequate visual perception ability of the model. 

\subsection{\errchip{errREASON}{Visual Reasoning Barriers}}

Visual reasoning barriers occur when a model accurately perceives relevant visual elements, such as colors, values, and spatial relationships, but fails to construct a correct logical inference from them. Unlike perception errors resulting from misunderstanding visual encoding, reasoning barriers reflect weaknesses in integrating perceptual input into a coherent conclusion. These errors frequently occur during tasks that require comparison, deduction, or multi-step reasoning. They may occur even when the initial visual reading is correct. Visual reasoning barriers may originate from earlier perceptual inaccuracies, as observed in multiple error patterns, but can also occur independently when the model commits to an interpretation or applies a flawed logical structure. Four major forms of visual reasoning barriers are identified: \textit{\textbf{R1 "Incorrect comparison"}} , \textit{\textbf{R2  "Incorrect reasoning logic"}},  \textit{\textbf{R3 "Perceptual–logic mismatch"}} and \textit{\textbf{R4 "Incomplete reasoning logic"}}.

% \begin{itemize}
% \item R1: Incorrect comparison
% \item R2: Incorrect reasoning logic
% \item R3: Perceptual–logic mismatch
% \item R4: Incomplete reasoning logic
% \end{itemize}

\noindent{\errchip{errREASON}{R1: Incorrect comparison}} Incorrect comparison errors occur when the model can identify relevant values or visual features but draws an incorrect comparative conclusion (e.g., concluding a category with a higher bar is smaller). In these instances, the perception ability remains valid, but the relational judgment between elements is incorrect. Such failures are common in comparison tasks, where the model must reason about relative magnitudes rather than simply retrieving values.

\noindent{\errchip{errREASON}{R2: Incorrect reasoning logic}}
Incorrect reasoning logic reflects a breakdown in the causal or inferential structure applied to the visual information. For example, a model may infer that \texttt{"A implies not B"} despite correctly recognizing that the visualization shows the opposite relationship. These errors result from insufficient logical deductions and often arise when the model introduces unsupported assumptions or applies improper reasoning to the visual content.

\noindent{\errchip{errREASON}{R3: Perceptual–logic mismatch}}
Perceptual–logic mismatch errors occur when perceptual input and logical inference diverge. The model may read values correctly, for instance, identifying \texttt{"A = 79 and B = 60"}, yet derive the conclusion that \texttt{"A is smaller than B."} This suggests that the model’s reasoning process may be subject to an internal heuristic overriding the perceptual evidence. Such mismatches highlight a disconnect between visual decoding and the reasoning process, similar to conflicts observed in human cognition.

\noindent{\errchip{errREASON}{R4: Incomplete reasoning logic}}
Incomplete reasoning logic occurs when the model reaches a conclusion without incorporating all relevant visual information or articulating intermediate reasoning steps. The error occurs due to a lack of critical comparisons in the model. The model could overlook global structure or fail to integrate multiple data components. Hence, the model "jumps to conclusions," producing answers that appear confident but lack sufficient analytic grounding.

\subsection{\errchip{errCOHERENCE}{Coherence Barriers}}

Coherence barriers arise when the model fails to maintain internal consistency throughout the reasoning process. Unlike visual reasoning barriers, which concern the correctness of inference based on perceived visual information, coherence barriers reflect disruptions in the model’s reasoning trajectory. These errors are often observed when the model generates intermediate reasoning steps that conflict with its final answer. Moreover, they occur  when external factors influence the answers, such as option ordering. Coherence barriers are well-recognized in MLLMs due to their generative process~\cite{wang2023selfconsistencyimproveschainthought, wei2023chainofthoughtpromptingelicitsreasoning}, in which later reasoning steps may overwrite or diverge from earlier ones. We identify two primary types of coherence barriers: \textit{\textbf{C1 "Self-consistency issues"}} and \textit{\textbf{C2  "Ordering affects answer"}}.

% \begin{itemize}
% \item C1: Self-consistency issues
% \item C2: Ordering affects answer
% \end{itemize}

\noindent{\errchip{errCOHERENCE}{C1: Self-consistency issues}}
Self-consistency issues occur when the model produces reasoning that contradicts its final answer. For example, the model may correctly infer in its intermediate reasoning that \texttt{"A is greater than B,”} yet output an answer indicating that B is larger. The inconsistency suggests that the reasoning chain and the decision output are not aligned, indicating a failure to maintain coherent internal logic. Such failures occur when the model revises or re-evaluates its intermediate interpretation, leading to a different conclusion than its initial statement. In some cases, the model tends to arrive at two conflicting internal claims, namely, \texttt{"A > B"}, coupled with \texttt{"B > A"}. This demonstrates instability in its reasoning process.

\noindent{\errchip{errCOHERENCE}{C2: Ordering affects answer}}
Ordering effects arise when the sequence in which information or answer choices are presented impacts the model’s response. In these cases, the model may respond correctly under one choice ordering but incorrectly under another, despite the  same underlying task and visual content. It implies that the model’s decision-making process is sensitive to coding structural variations rather than grounded in visual reasoning. As such, ordering effects suggest that MLLMs may rely on positional heuristics or latent biases toward earlier answer options, rather than grounding their reasoning within the visualization. 

\begin{tcolorbox}[enhanced jigsaw, breakable, colback=gray!5, colframe=black!60,
  boxrule=0.5pt, arc=2pt, left=4pt, right=4pt, top=4pt, bottom=4pt,
  fontupper=\footnotesize\ttfamily, title=Open Coding Example – Bar Chart Comparison Error,
  sharp corners=south, listing only,
  listing options={basicstyle=\ttfamily\footnotesize, breaklines=true, breakatwhitespace=true}
]\label{bar_comparision_error}

\textbf{Case ID:} bar\_chart\_Q4\_\emph{Claude}\newline
\textbf{Question:} How many countries in Asia have an average Internet speed slower than South Korea?\newline
\textbf{Type:} Make Comparisons (absolute value)\newline
\textbf{Correct Answer:} 8 countries\newline
\textbf{Model Answer:} 11 countries\newline

\textbf{Model Reasoning Trace:}

"I need to identify South Korea's internet speed and count how many countries
have slower speeds. Looking at the chart, South Korea appears to have an 
internet speed of approximately 14.5 Mbps. Examining each bar, I can see that 
the following countries have speeds below South Korea's: China (16.8), 
Hong Kong (12), India (9.8), Indonesia (13.2), Japan (11.8), 
New Zealand (10.2), Singapore (12.8), Taiwan (13.2), Thailand (10), 
and Vietnam (16.8). Wait, let me recheck – South Korea actually appears to 
be around 14.5 Mbps, and counting countries below this level, I count 
11 countries with slower speeds."\newline

\textbf{Model Coordinates:}\newline
(693, 400): "South Korea’s bar – estimate speed value",\\
(400, 300): "Scanning across all bars for comparison",\\
(200, 450): "Countries with lower bars",\\
(800, 350): "Middle-right group of bars",\\
(600, 400): "Relative heights near South Korea".\newline

% \textbf{Primary Barrier Codes (identified during open coding):}
% \begin{itemize}[leftmargin=*, itemsep=1pt, label=--]
%   \item R1: Incorrect comparison – incorrectly comparing to the bar heights and hence confuses taller bars as slower speeds.
%   \item R4: Incomplete reasoning logic – inability to complete comparison logic with the right conclusion.
%   \item C1: Self-inconsistency – reasoning shifts mid-sentence (“Wait, let me recheck”).
% \end{itemize}

% \textbf{Evidence Quotes:}
% \begin{itemize}[leftmargin=*, itemsep=1pt, label=--]
%   \item “Looking at the chart, South Korea appears to have an internet speed of approximately 14.5 Mbps.” – initial misread of baseline.
%   \item “China (~16.8)” and “Vietnam (~16.8)” – contradictory inclusion of higher-valued countries as “slower.”
%   \item “Wait, let me recheck …” – self-correction attempt revealing confusion about axis meaning.
% \end{itemize}

% \textbf{Barrier Chain:}\newline
% R1 - R4 - C4 → Incorrect Answer\newline
\end{tcolorbox}

% To demonstrate the taxonomy in practice, we use a representative example from \emph{Claude} shown in Example~\ref{bar_comparision_error}. The question deals with a bar chart \texttt{"How many bars are lower than the highlighted bar?"} The correct answer, determined by manual counting, is 8 countries, but the model mistakenly reports 11. The model correctly reads 10 country values but incorrectly compares them to the highlighted bar. This response demonstrates both~\errchip{errREASON}{visual reasoning barriers}, such as selecting the wrong extrema due to local focus and~\errchip{errCOHERENCE}{coherence barriers}, like providing incomplete or logically inconsistent justifications. Illustrated in Figure~\ref{fig:bar-q4-incorrect}, the barriers are classified as a combinations of~\errchip{errREASON}{R1: incorrect comparison, R4: incomplete reasoning logic} and~\errchip{errCOHERENCE}{C1: Self-consistency issues}. These findings indicate that MLLMs can localize focus accurately but often struggle with visual reasoning when presented with many objects, leading to incorrect object counting. 
\noindent To demonstrate the taxonomy in practice, we present a representative example from \emph{Claude} depicted in Example~\ref{bar_comparision_error}. The question asks: \texttt{"How many bars are lower than the highlighted bar?"} The correct answer is 8 countries while the model reports 11. The answer exemplifies both~\errchip{errREASON}{Visual Reasoning Barriers} and~\errchip{errCOHERENCE}{Coherence Barriers}. As shown in Figure~\ref{fig:bar-q4-incorrect}, the model exhibits~\errchip{errREASON}{R1: Incorrect comparison} of mapping bar height to slower speeds: \texttt{"Looking at the chart, South Korea appears to have an internet speed of approximately 14.5 Mbps."} The model also demonstrates~\errchip{errREASON}{R4: Incomplete reasoning logic}, for example, stating China (16.8) and Vietnam (16.8) as \texttt{"slower,"} revealing inability to reach the correct comparison. Additionally,~\errchip{errCOHERENCE}{C1: Self-consistency issues} arise when the model shifts states \texttt{"Wait, let me recheck,"} which indicates decision instability. This case thus illustrates that, although MLLMs can visually localize chart elements, they struggle with accurate comparative reasoning across multiple objects, leading to incorrect object counting.

\begin{figure}[h]
  \centering
  % Adjust width as needed: 0.9\linewidth for one-column, 0.48\linewidth for two-column
  \includegraphics[width=\linewidth]{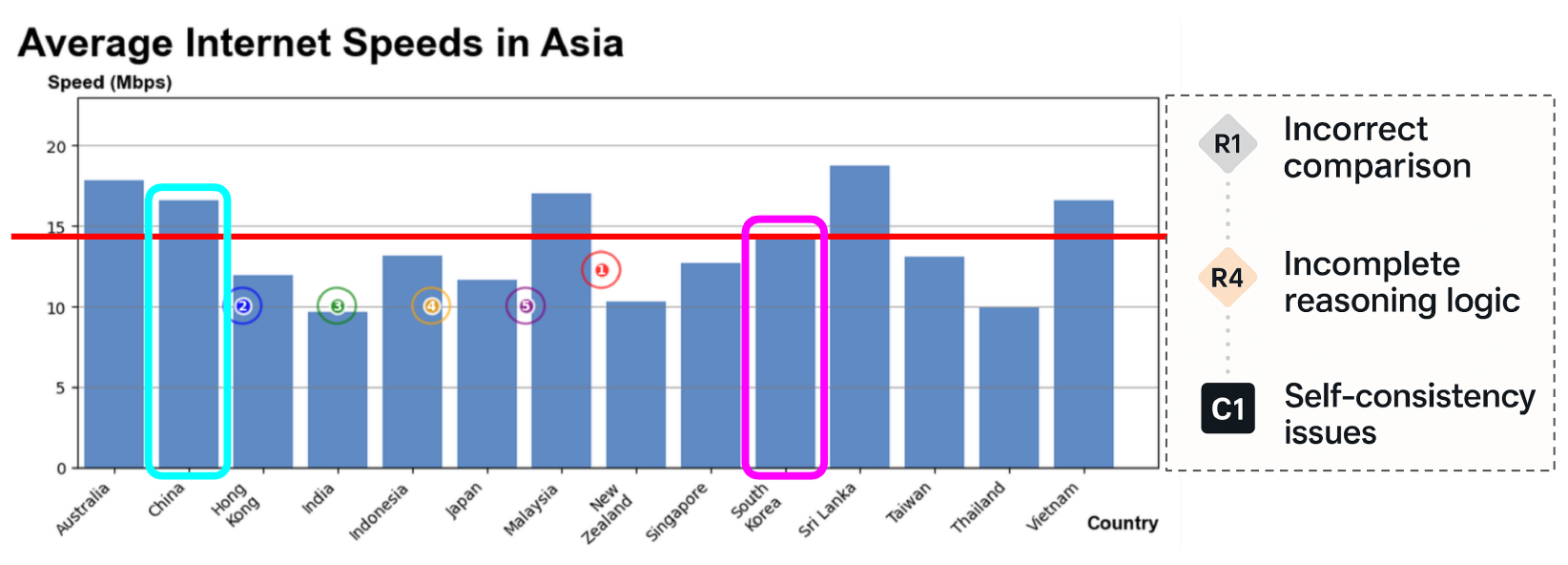}
  \caption{
      An example of an incorrect MLLM response on a bar chart task (Q4). Highlighted regions denote the model’s reported attention points and 
      their misidentified maxima. The red horizontal line shows the threshold for correctly identifying countries with speeds above 15 Mbps.
  }
  \label{fig:bar-q4-incorrect}
\end{figure}

\section{Quantitative Analysis}
In this section, we first summarize overall model performance, both in terms of accuracy and by chart type. We then examine error patterns through a barrier analysis, using the taxonomy introduced in Section~\ref{sec:barrier} and illustrating key failures with example responses from \emph{Claude}.

\subsection{Overall performances}

\noindent{\textbf{Performance accuracy.}}
% 4 numbers
Across the 53 reVLAT questions and 5 randomized orderings per question, the accuracies of 4 models vary. \emph{Qwen} achieves the strongest overall performance at 86.42\%, followed by \emph{Claude} at 74.72\%, while \emph{GPT} at 49.06\% and \emph{Gemini} at 73.21\% show more moderate accuracy with higher variability across orderings. \newline

\noindent{\textbf{Performances by chart type.}}
Figure~\ref{fig:accuracy-chart-type} breaks down accuracy across the 12 chart types in reVLAT, revealing several consistent cross-model patterns. 
% First, despite overall gains compared to prior VLAT-style evaluations~\ref{}, where earlier MLLMs typically scored in the \% range. It demonstrates that modern models from literature still exhibit weaknesses in specific visualization categories~\ref{}. 
\emph{Claude} achieves strong performance on common chart types such as bar charts and histograms, but its accuracy drops on pie charts (46.70\%) and stacked bar charts (72.00\%), where nearly half of its total errors occur. \emph{Qwen}, the best-performing model overall, follows a similar trend: while it attains 100\% accuracy on area charts, choropleth map, histograms, etc., its performance noticeably decreases on pie charts (66.70\%) and stacked bar charts (52.00\%).

\noindent A pattern emerges in Figure~\ref{fig:accuracy-chart-type} that charts that require interpreting multiple colored segments, such as pie charts and stacked bar charts, consistently lead to lower accuracy across all four models. Our observations align with prior findings that MLLMs struggle when categorical color encodings interact with quantitative magnitude estimation. In these charts, models must simultaneously parse encoding for category identification and angle/height/area for value estimation. We hypothesize that this dual dependence on color and magnitude introduces additional complexity in the visual embedding space, making quantitative judgments less stable and more prone to error. In addition, these chart types also appear less frequently in current pretraining corpora, which may further limit the models' familiarity with the visual patterns required to answer such questions reliably.

\noindent \textbf{Insights.} The chart-type results indicate that (1) modern MLLMs have made progress on visualization literacy compared to past work, (2) errors are no longer uniformly distributed but concentrated in color-heavy, segment-based chart types, and (3) even high-performing models such as \emph{Qwen} remain vulnerable to misinterpretations when color and quantity must be jointly decoded. These findings reinforce the need to evaluate MLLMs not solely on overall accuracy, but on their robustness across diverse visual encoding and design choices.% \hl{ordering consistent with accuracy bar}
\begin{figure}[t]
    \centering
    \includegraphics[width=0.9\columnwidth]{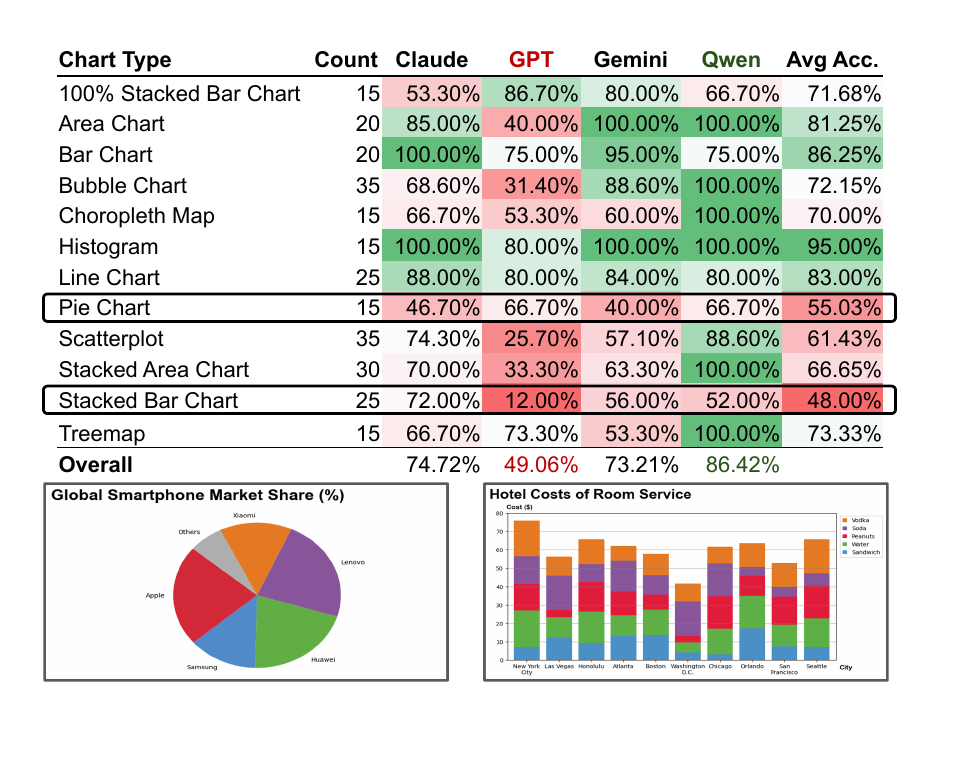}
    \caption{
        Accuracy by chart type across 4 MLLMs. Cells are colored from red (lowest accuracy) to green (highest). This visualization highlights per-chart difficulty and model-specific weaknesses.
    }
    \label{fig:accuracy-chart-type}
\end{figure}

% \begin{table}[h!]
% \centering
% \small
% \begin{tabular}{lcccc}
% \toprule
% \textbf{Chart Type} & \textbf{Claude-Sonnet-4} & \textbf{OpenAI} & \textbf{Gemini} & \textbf{Qwen3-VL} \\
% \midrule
% 100\% Stacked Bar Chart & 89.6\% & 98.5\% & 95.8\% & 86.1\% \\
% Area Chart              & 95.5\% & 91.1\% & 100.0\% & 100.0\% \\
% Bar Chart               & 100.0\% & 96.3\% & 98.6\% & 86.1\% \\
% Bubble Chart            & 83.6\% & 82.2\% & 94.4\% & 100.0\% \\
% Choropleth Map          & 92.5\% & 94.8\% & 91.5\% & 100.0\% \\
% Histogram               & 100.0\% & 97.8\% & 100.0\% & 100.0\% \\
% Line Chart              & 95.5\% & 96.3\% & 94.4\% & 86.1\% \\
% Pie Chart               & 88.1\% & 96.3\% & 87.3\% & 86.1\% \\
% Scatterplot             & 86.6\% & 80.7\% & 78.9\% & 88.9\% \\
% Stacked Area Chart      & 86.6\% & 85.2\% & 84.5\% & 100.0\% \\
% Stacked Bar Chart       & 89.6\% & 83.7\% & 84.5\% & 66.7\% \\
% Treemap                 & 92.5\% & 97.0\% & 90.1\% & 100.0\% \\
% \midrule
% \textbf{Overall}   & \textbf{91.68\%} & \textbf{91.66\%} & \textbf{91.67\%} & \textbf{91.67\%} \\
% \bottomrule
% \end{tabular}
% \caption{Accuracy of Models Across Chart Types}
% \label{tab:model_chart_accuracy}
% \end{table}

% \textbf{Insights} some models good or bad
\subsection{Barrier Analysis across models}
In total, we have open-coded 309 erroneous responses across models, including 67 by \emph{Claude}, 135 by \emph{GPT}, 71 by \emph{Gemini}, and 36 by \emph{Qwen}.

\noindent{\textbf{\emph{Claude}.}} Figure~\ref{fig:error-distribution-claude} presents the distribution of barrier types for the 67 erroneous responses from \emph{Claude} across all chart types. Charts with richer color encodings, particularly stacked bar and pie charts, reveal a consistent pattern that~\errchip{errPERPCEPTION}{V2 Misreading values} accounts for a substantial proportion of errors. The barrier dominates the error distributions for these chart types. This pattern suggests that the presence of multiple colored segments and legends impedes the model’s ability to interpret or estimate numeric values when spatial attention is well aligned. We hypothesize that introducing multiple shades into the visual embedding space increases representational complexity. Hence, it challenges the model to distinguish color as a categorical cue from quantitative encodings such as height, angle, or area. In contrast, for visually simpler charts, such as line and area charts, errors are more evenly distributed across~\errchip{errREASON}{Visual Reasoning Barriers} and~\errchip{errCOHERENCE}{Coherence Barriers}. This indicates that value misinterpretation is a sensitive failure mode when color and quantity must be decoded together from the visualization.
\begin{figure}[t]
    \centering
    \includegraphics[width=\columnwidth]{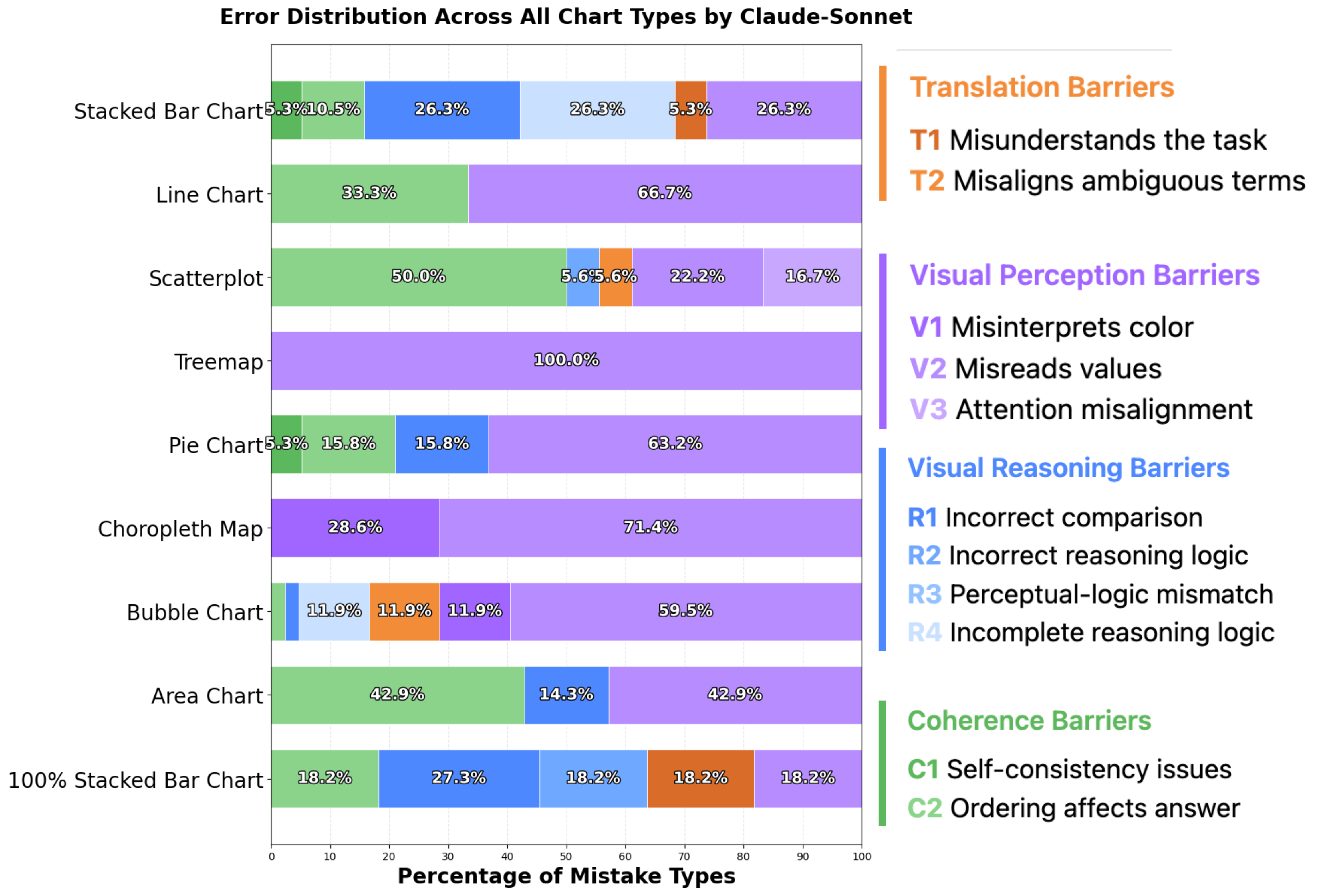}%
    \caption{Error distribution across all chart types for \emph{Claude}. 
    Bars show the percentage of each mistake type per chart, grouped across taxonomy described in Section~\ref{sec:barrier}. }
    % \textbf{\errchip{errTrans}{Translation barriers}} (T1: Misunderstands the task, T2: Misaligns ambiguous terms), 
    % \textbf{\errchip{errPERPCEPTION}{Visual perception barriers}} (V1: Misinterprets color, V2: Misreads values, V3: Attention misalignment), 
    % \textbfA{\errchip{errREASON}{Visual reasoning barriers}} (R1: Incorrect comparison, R2: Incorrect reasoning logic, 
    % R3: Perceptual-logic mismatch, R4: Incomplete reasoning logic), and 
    % \textbf{\errchip{errCOHERENCE}{Coherence barriers}} (C1: Self-consistency issues, C2: Ordering affects answer).}
    \label{fig:error-distribution-claude}
\end{figure}

\section{Limitations and Future Directions}
% \paragraph{Limitations and risks}
Several limitations should be acknowledged. First, MLLMs are sensitive to prompt design, and our use of a minimal prompting strategy may not fully represent performance in real-world application scenarios. Future work could employ more sophisticated prompting techniques to replicate our analysis and potentially reveal different failure patterns. Second, the SOTA MLLMs faces challenges to natively sketch or annotate visualizations during inference. While we address the limitation by requesting coordinate outputs to capture spatial attention textually, our framework can be extended to accommodate future MLLMs with multimodal output capabilities.

In the future, we will focus on expanding the breadth and depth of our analysis across models and task conditions. First, scaling up the ordering manipulations will enable us to collect richer behavioral data and more robustly characterize the extent to which presentation structure influences MLLM responses. Second, we aim to automate portions of the closed coding process using LLM-as-judge approaches, enabling more efficient, replicable annotation at scale while preserving interpretive rigor~\cite{zheng2023judgingllmasajudgemtbenchchatbot}. Finally, we plan to design targeted experimental interventions to directly test the hypothesized mechanisms behind each barrier class, providing stronger causal evidence for why MLLMs fail at specific stages of visualization interpretation and reasoning.

\section{Conclusion}
In this work, we investigated whether MLLMs see visualizations in ways comparable to humans, and why they fail when interpreting chart types. By adapting the barrier-centric methodology of~\cite{nobre2024reading} to AI systems, we developed the first systematic taxonomy of visualization literacy barriers for MLLMs, derived through open coding of 309 erroneous responses across 4 SOTA models. Our taxonomy extends prior human-focused frameworks by identifying two machine-specific barrier classes, namely~\errchip{errREASON}{Visual Reasoning Barriers} and~\errchip{errCOHERENCE}{Coherence Barriers}.

Quantitative analyses reveal that errors are not uniformly distributed: models perform well on visually simple charts but struggle on color-intensive, segment-based visualizations, where categorical encoding and magnitude estimation must be jointly resolved. Barrier distributions further show that even when MLLMs accurately localize relevant visual elements, they frequently fail to construct consistent comparative reasoning, leading to incorrect conclusions.

These findings provide new insights into the cognitive limitations of MLLMs in visualization interpretation, with implications for their deployment as analytical assistants. Understanding where and why MLLMs fail offers a foundation for improving model design, evaluation protocols, and human–AI collaboration in visual analysis contexts. Looking forward, scaling behavioral sampling, automating coding, and designing targeted experimental ablations will enable stronger causal evidence for the mechanisms underlying these barriers. Together, our study aims to assist in establishing guidelines for more reliable and trustworthy use of MLLMs’ visual reasoning.

\acknowledgments{
The authors gratefully acknowledge the teaching support and guidance provided by Matthew Varona and Prof. Carolina Nobre.
}

\bibliographystyle{abbrv-doi-hyperref}
%\bibliographystyle{abbrv-doi-hyperref-narrow}
%\bibliographystyle{abbrv-doi}
%\bibliographystyle{abbrv-doi-narrow}

% \bibliography{template}

\bibliography{reference}
\appendix % You can use the `hideappendix` class option to skip everything after \appendix
% \section{VLAT dataset regeneration}

% \section{Open encoding}

\section*{An example of system prompt demonstration} \label{sec:prompts_system}
\begin{tcolorbox}[enhanced jigsaw, breakable, colback=gray!5, colframe=black!60,
  boxrule=0.5pt, arc=2pt, left=4pt, right=4pt, top=4pt, bottom=4pt,
  fontupper=\footnotesize\ttfamily, title=System Prompt,
  sharp corners=south, listing only,
  listing options={basicstyle=\ttfamily\footnotesize, breaklines=true, breakatwhitespace=true}
]
You are an expert data visualization analyst. You will be shown a visualization and asked a question about it.\newline

Please provide your response in the following JSON format:\newline

\{
  "choice": "<the letter of your answer choice, e.g., 'a', 'b', 'c', etc.>",\newline
  "answer\_value": "<the actual text of the answer you selected>",\newline
  "rationale": "<your reasoning for selecting this answer in 5 sentences or less>",\newline
  "attention\_regions": [\newline
    \{"x": <x\_coordinate>, "y": <y\_coordinate>, "description": "<what you looked at>"\}\newline
  ]\newline
\}\newline

\textbf{Important Instructions:}\newline
1. The visualization image size is 1233 × 600 pixels.\newline
2. For \texttt{attention\_regions}, provide up to 5 coordinate points (x, y) where you focused when answering.\newline
3. \textbf{x-range:} 0 (left) → 1233 (right); \textbf{y-range:} 0 (top) → 600 (bottom).\newline
4. Each region should include a brief \textbf{description} of what you were looking at.\newline
5. Keep your \texttt{rationale} concise — 5 sentences or fewer.\newline
6. If unsure, select the \textbf{"Not sure"} option and explain why in your rationale.\newline
7. Always return \textbf{valid JSON only} (no Markdown or extra commentary).\newline
\end{tcolorbox}

\end{document}